# Statistical imaging of NV centers reveals clustered defect formation in diamond

**Jason Shao[1], Richard Monge[1], Tom Delord[1,†], Carlos A. Meriles[1,2,†]**

[1]*Department of Physics, CUNY- The City College of New York, New York, NY 10031, USA.*

[2]*CUNY-Graduate Center, New York, NY 10016, USA.*

[†]E-mails: tdelord@ccny.cuny.edu, cmeriles@ccny.cuny.edu.

The sharp optical resonances of $NV^-$ centers in diamond at cryogenic temperatures offer powerful new capabilities for material characterization, but extracting the most detailed information typically requires careful calibration of individual sensors, limiting scalability. In this work, we use resonant photoluminescence excitation imaging to optically resolve and monitor hundreds of individual NVs across large fields of view, enabling statistical analysis of their spatial distribution with sub-diffraction resolution. This multiplexed, non-destructive approach allows quantum sensors to characterize the material platform they inhabit. Focusing on CVD-grown diamond, we uncover significant deviations from random distributions, including an unexpectedly high occurrence of closely spaced clusters comprising two or more NVs. These findings suggest non-Poissonian formation dynamics and point to spatially correlated defect generation mechanisms. Beyond offering insight into diamond growth and NV center formation, our approach enables the scalable identification of naturally occurring NV clusters — configurations that are promising for entanglement-assisted quantum information protocols and correlated sensing — and establishes a path toward structural and electronic defect analysis in various material hosts at the single-emitter level.

## 1. INTRODUCTION

Traditional studies of color centers in solids have largely relied on confocal microscopy to investigate individual emitters, enabling high-resolution imaging, spectral characterization, and spin-based sensing with nanoscale precision[1,2]. While this approach has been instrumental in establishing color centers in diamond as leading candidates for quantum information and sensing applications, it is inherently limited by its serial nature. Scaling beyond isolated emitters — either to enable statistical population-level studies or to identify rare but technologically valuable configurations such as spin-coupled clusters — requires a shift toward parallelized strategies capable of resolving many color centers simultaneously.

Recent advances in wide-field and spectrally selective imaging have begun to make such transitions possible, opening new opportunities to characterize the properties of solid-state quantum systems at scale[3-8]. By combining spatial and spectral resolution, these approaches not only accelerate data acquisition but also enable entirely new types of analysis — in particular, statistical inference of the spatial and spectral distributions of embedded quantum defects[9-14]. In this context, the population of color centers itself becomes an object of study, revealing information about the material platform they inhabit. Questions of interest include whether defects form independently or through correlated processes[15], how growth conditions shape local impurity distributions[16], and whether spatial clustering reflects stochastic fluctuations or underlying physical mechanisms.

Here, we implement cryogenic photoluminescence excitation (PLE) imaging as a multiplexed platform to optically address hundreds of individual nitrogen-vacancy (NV) centers in diamond[17] with sub-diffraction spatial resolution. Focusing on crystals grown via chemical vapor deposition (CVD), we use this method to carry out statistical mapping of NV center positions across large areas. Our analysis reveals significant deviations from a Poissonian distribution, with an overrepresentation of closely spaced NV pairs and higher-order clusters. These findings suggest spatially correlated formation dynamics and point to underlying defect interactions or inhomogeneities during growth. Beyond their relevance for materials science, these naturally formed NV clusters represent a previously untapped quantum resource, with potential applications in entanglement-enabled sensing, multi-qubit logic, and spin-based quantum memory. In addition to providing a framework for investigating emergent quantum structure in solid-state systems, our analysis methods are broadly applicable to correlated sensing scenarios[4-6], including structural mapping of charge traps at the single-defect level[7,8].

## 2. RESULTS

Our experimental platform is designed to perform resonant photoluminescence excitation (PLE) imaging of individual NV centers under cryogenic conditions[18]. At low temperatures, the negatively charged NV exhibits a collection of narrow optical transitions near 637 nm, with typical inhomogeneous linewidths on the order of ~100 MHz[19]. These resonances, arising from spin-conserving transitions between the ground and excited triplet states, vary from one NV to another — even for centers spaced only nanometers apart — due to local variations in electric and strain fields within the crystal[20-22]. By addressing these transitions with a narrowband, frequency-tunable laser, we achieve spectral selectivity sufficient to isolate and characterize individual NVs within dense ensembles.

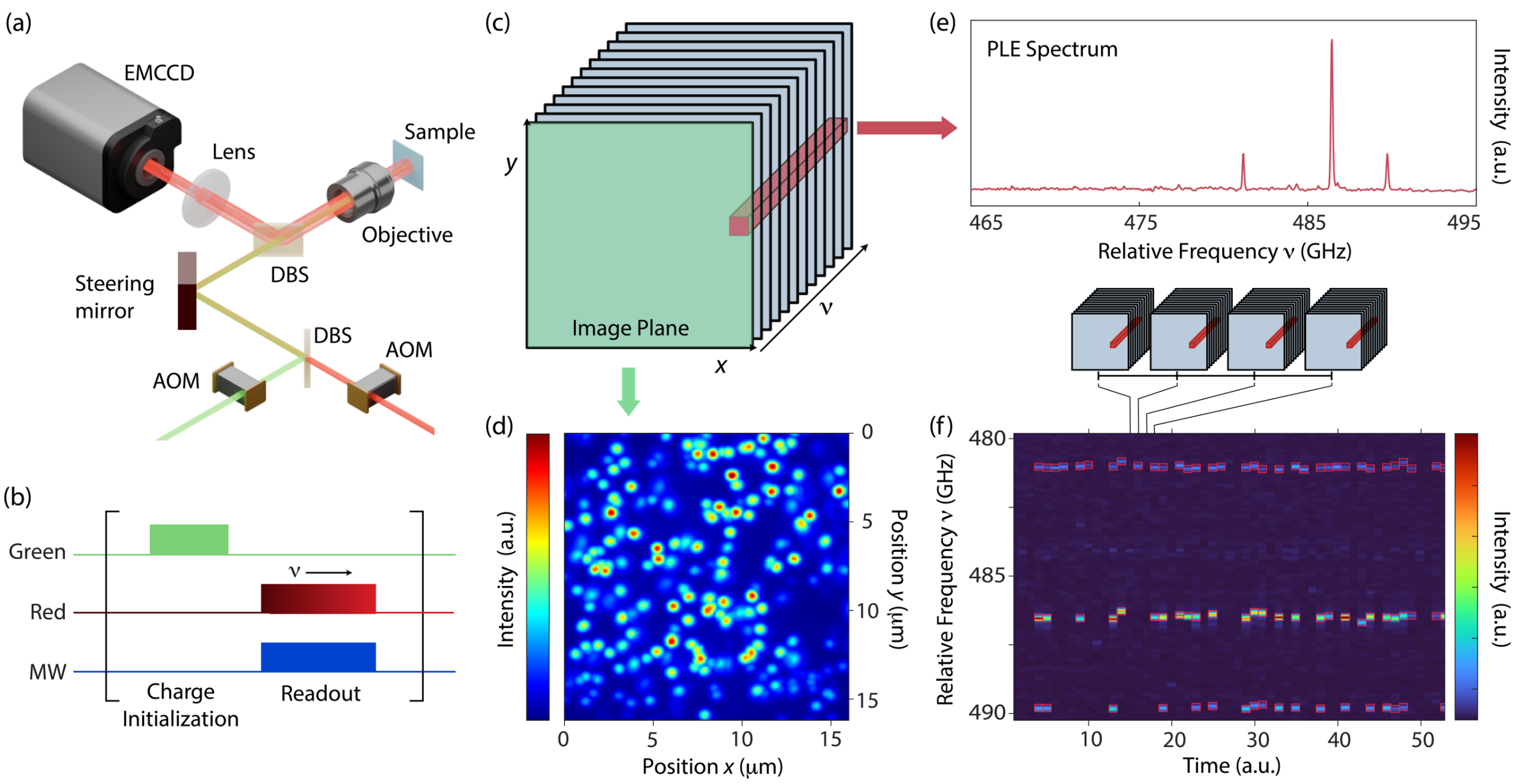

**Fig. 1: Wide-field PLE setup and dataset structure.** (a) Schematic of the experimental setup. The microscope uses a green (532 nm) and a red (637 nm) narrowband tunable laser, each controlled via individual acoustic optical modulators (AOM). Dichroic beam splitters (DBS) separate or combine the beams, and images are captured by an EMCCD camera. (b) PLE spectroscopy protocol. Following NV charge state initialization via widefield green excitation, the EMCCD camera records widefield images of the entire NV set as the red laser sweeps a frequency range sufficiently broad to encompass the optical transitions of the NVs in the set; continuous MW excitation resonant with the NV ground state crystal field splitting prevents NV spin population trapping away from optical resonance. (c) Visual representation of the resulting 3D dataset, comprising 2D images of the field of view at each of the red laser frequencies. (d) Projection of the data stack along the frequency axis; the result is a 2D map of the full NV set. (e) PLE spectrum as derived from the PL amplitude recorded by the EMCCD camera at a given position. (f) Multiple repetitions of the same protocol at successive times results in a 4D dataset. The plot shows a time series of the PLE spectrum in (e). Vanishing resonances indicate improper charge initialization or ionization during the laser scan. Experiments are carried out at 7 K.

Figure 1a shows a schematic of the wide-field imaging setup. A green (532 nm) laser initializes NVs into the negative charge state, while a red (637 nm) tunable laser scans across the zero-phonon line (ZPL) to probe optical resonances. Each beam is modulated via its own acousto-optic modulator (AOM), and the beams are combined or separated using dichroic optics. Fluorescence is collected through a high-numerical-aperture objective and imaged onto an EMCCD camera, enabling parallel acquisition of photoluminescence (PL) signals from all emitters within the field of view[23] (see Appendix A)

The spectroscopy protocol, illustrated in Fig. 1b, begins with widefield green laser illumination to initialize the charge state across the ensemble preferentially into negative. The red laser then sweeps a frequency range sufficiently broad (~30 GHz) to encompass the optical resonances of the NVs in the region. To prevent optical pumping into dark spin states, we apply a continuous microwave field resonant with the NV ground-state zero-field splitting throughout the scan. As the red laser tunes across optical resonances, the EMCCD camera records a stack of wide-field PL images (Fig. 1c), forming a three-dimensional dataset with two spatial dimensions and one spectral axis.

Calculating the photoluminescence (PL) maximum across the frequency dimension for each pixel yields a two-dimensional (2D) image of the NV ensemble (Fig. 1d). At each pixel, the PL intensity as a function of excitation frequency provides a PLE spectrum (Fig. 1e), from which we extract the resonance frequencies and optical linewidth. We typically repeat this protocol multiple times (50 or more), each time using widefield green illumination to reset NV centers into their negative charge state. The resulting dataset can therefore be seen as a 4D scalar field comprising spatial, spectral, and temporal information, which allows us to track spectral fluctuations and charge stability at the single-emitter level (Fig. 1f). Transient disappearance of resonances indicates failed charge initialization or photoionization during the scan. This latter possibility, however, can be minimized thanks to the low power (μW-nW) typical in our PLE experiments, effectively rendering optical sweeps non-destructive[22]. This approach enables the high-throughput optical characterization of hundreds of NVs in parallel with sub-diffraction resolution, thus combining spectral precision with spatially resolved access.

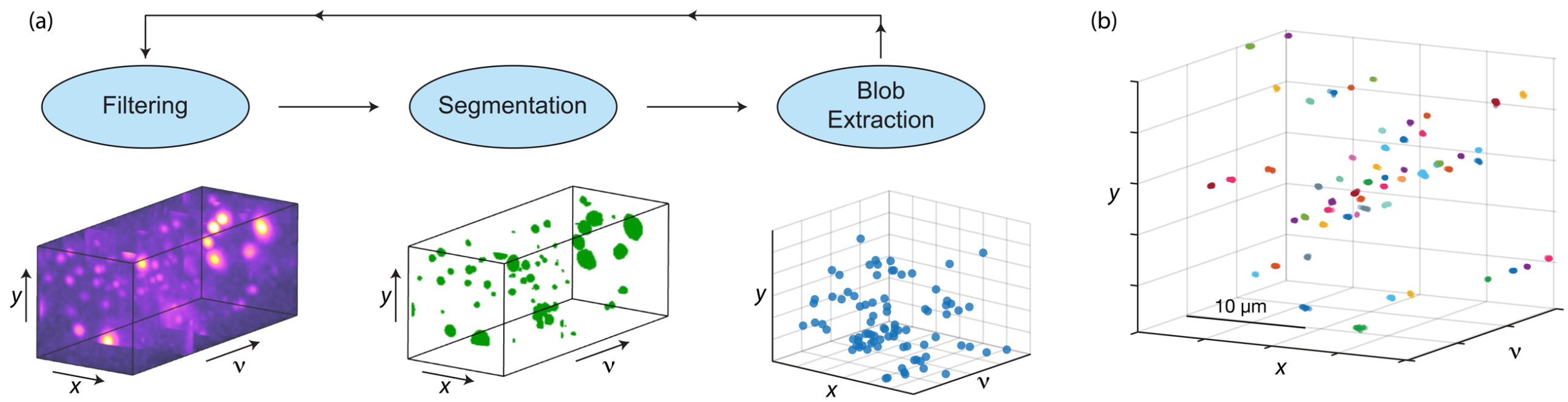

**Fig. 2: Data processing and feature extraction.** (a) Procedure to extract blobs from each spectrum, along with visuals of the results in each step, namely, filtered, binarized, and scatter plot data of characterized blobs as a function of position and frequency. (b) Blobs accumulated across all spectra, visualized and categorized into corresponding optical transitions by their positions in real space and frequency. Colors differentiate nearby blobs corresponding to different optical resonances.

Following raw data acquisition, we apply a multi-step processing pipeline to enhance signal-to-noise ratio (SNR) and extract relevant spectral features (Fig. 2). Each spectrum is first smoothed using a 3D Gaussian filter applied in both spatial and frequency dimensions, with kernel widths chosen to match the expected signal profile — 400 nm in real space and 100 MHz in frequency. The filtered data is then binarized using an adaptive thresholding algorithm that mitigates the influence of local photoluminescence background. In the resulting binary map, optical resonances appear as continuous bright regions or "blobs," which are identified and characterized using a custom-made blob detection and processing algorithm.

To suppress false positives due to noise or background fluctuations, detected blobs are filtered based on morphological and spectral parameters (e.g., volume, intensity). The outcome is a curated list of optical resonances, each defined by its spatial position, central frequency, and linewidth. These parameters can be further refined by fitting the original (unfiltered) data with a Gaussian model. Data analysis is then performed across multiple spectra, leveraging repeated measurements to improve positional precision and to reliably associate optical resonances with individual NV centers.

To illustrate our resonance grouping approach, Fig. 3 presents a representative case in which six optical resonances are detected within a localized region of the sample. The confocal image in Fig. 3a shows a small group of NVs, including two prominent bright spots. The transitions we focus on in this analysis originate from the dimmer region, where multiple resonances cluster. These transitions, not resolved spatially in the confocal image, are identified and localized through Gaussian fitting to a set of EMCCD images acquired at different laser frequencies and over several repetitions of the frequency sweep. Fig. 3b shows the outcome of this analysis: Each peak represents the spatial distribution of a single optical resonance reconstructed from the fitted EMCCD spot positions: The lateral separation between clusters suggests the presence of more than one emitter. To further support this assignment, we examine the ionization dynamics of each resonance across time (Fig. 3c). Resonances belonging to the same NV tend to ionize and recover in synchrony, yielding similar blinking trajectories (even though weaker resonances tend to appear more irregularly due to failed identification by the automated thresholding software, see traces for resonances v and vi).

We assess the likelihood that any two detected optical resonances originate from the same NV center by integrating both positional and temporal information into a unified probabilistic framework. Specifically, we compute Bayes factors that compare the probability of a single-emitter hypothesis against that of two distinct emitters. For the spatial component, this factor incorporates the relative displacement between resonances, the measurement uncertainty, and the local NV density. For the temporal component, we evaluate the synchronicity of their blinking behavior using both positive and negative correlations, weighted by each resonance's detection probability across time. These complementary metrics (quantified in Figs. 3d and 3e; see also SOM) are then combined multiplicatively to produce a composite association score for each pair. The resulting matrix reveals a block-diagonal structure consistent with two distinct NV centers. This integrated approach provides a robust means of resolving closely spaced emitters in regimes where spatial overlap or spectral crowding alone would obscure their identity (see Appendix B for details).

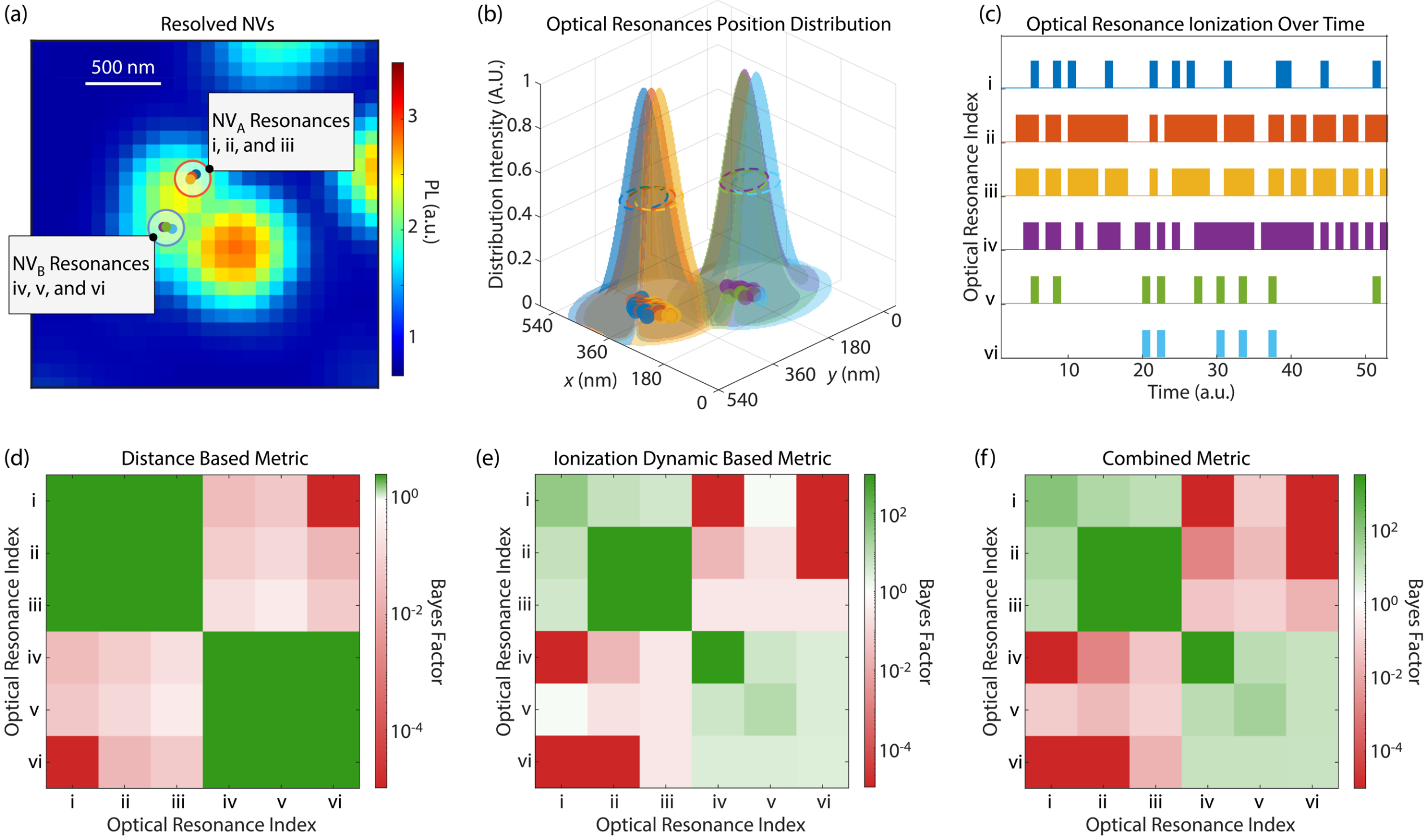

**Fig. 3: Data processing and identification of individual NVs.** (a) Confocal image showing a region with multiple NV centers; the six resonances analyzed here originate from the dimmer feature near the center. (b) Spatial distribution of resonance positions, extracted from EMCCD data and fit with Gaussian profiles; each peak corresponds to a single optical transition, with height reflecting relative localization confidence. (c) Ionization time traces for the six resonances. (d-f) Likelihood that both resonance originates from the same emitters over likelihood they originate from different emitters (Bayes factor) using positional information (d), ionization dynamic (e) and combined information (f).

To quantitatively assess the spatial organization of NV centers, we performed statistical analysis on two CVD electronic-grade diamond samples (Diamonds A and B) where NV formation was stimulated by implantation damage and subsequent annealing[24,25]. Note that we specifically analyze non-implanted regions, towards which vacancies migrated. As shown in Fig. 4, we identify NV locations (white crosses) from the processed frequency-integrated PLE maps and assign clusters (red circles) to NVs that fall within the optical diffraction limit of their nearest neighbor. In both samples, a substantial fraction of NVs appear in close proximity, forming distinct spatial groupings rather than being uniformly distributed across the field of view. The persistence of this clustering behavior across independently prepared samples suggests that it reflects an intrinsic feature of the post-annealing defect distribution.

To compare this observed clustering behavior with a null hypothesis of spatial randomness, we calculated the cluster size distribution, defined as the probability of finding an "*n*-cluster" containing exactly *n* NVs (Fig. 4b). The experimental distribution (light blue bars) reveals a pronounced enhancement in the formation of multi-NV clusters compared to a synthetic dataset generated under the assumption of complete spatial randomness (CSR) with the same areal density (light red bars). This excess probability for larger clusters suggests an intrinsic correlation in NV positioning. Further, upon quantifying the degree and scale of clustering using Ripley's *K*-function — a standard measure of spatial point–pattern correlation, not shown — we find values consistently above the CSR expectation across a broad range of radii. Importantly, the deviation from randomness is strongest below ~2 μm and peaks near ~100 nm, indicating that NV centers are markedly more likely to occur in close proximity than a Poisson model would predict.

It is worth noting that because wide-field imaging provides high lateral localization precision but limited axial resolution, the spatial coordinates extracted from the PLE maps correspond to a projection of the three-dimensional NV distribution within a finite axial slice (~3 μm) onto the imaging plane. This sampling introduces two opposing effects[26]. On the one hand, projection can bring emitters that are separated along the optical axis into apparent lateral proximity, slightly redistributing correlations within the observed slice. On the other hand, NV centers located just outside the observed axial region are not detected, which tends to reduce the apparent clustering (a boundary effect). These effects are expected to be small under our experimental conditions and are accounted for in the comparison with synthetic datasets, which are constructed using the same projected density. Consequently, projection does not artificially produce the observed excess of closely spaced NV pairs but would instead tend to distort or dilute genuine correlations present in the underlying three-dimensional distribution.

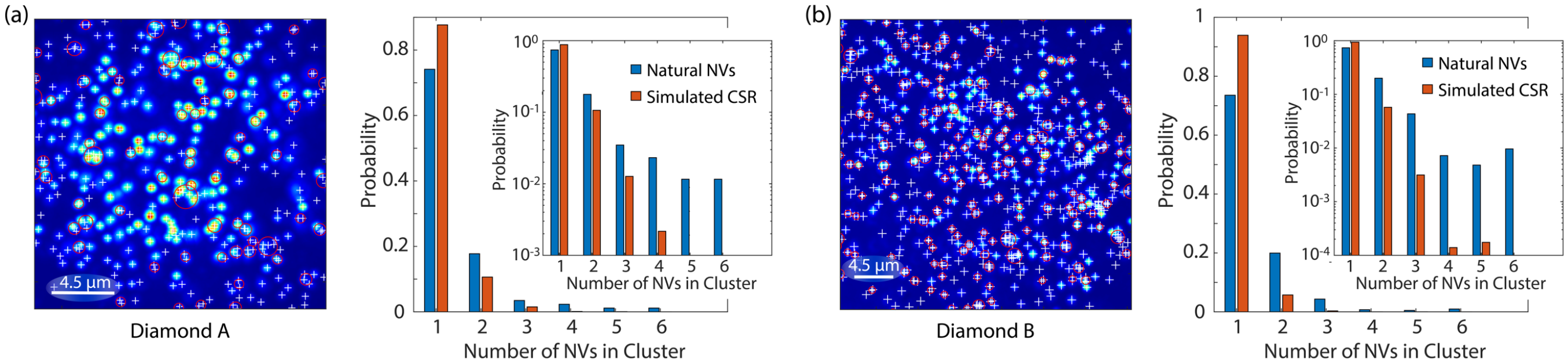

**Fig. 4: Statistics of NV cluster formation.** (a) (Left) Frequency-integrated PLE map of diamond A. After processing, we find that a large fraction of the identified NVs (white crosses) fall within diffraction distance from their closest neighbors to form "clusters" (red circles with size proportional to the number of NVs). (Right) Probability distribution for the formation of clusters containing *n* NVs (i.e., an "*n*-cluster", light blue bars). Compared to the fractions derived from assuming the same areal concentration but complete spatial randomness (CSR, red bars), naturally occurring NVs are more likely to group. (b) Same as in (a) but for diamond B.

The microscopic origin of the observed NV clustering remains an open question. Since NV centers form during annealing via diffusion and capture of vacancies at substitutional nitrogen sites, the spatial correlations we observe could reflect either pre-existing inhomogeneities in nitrogen incorporation or biases in the vacancy diffusion process during annealing. In the former scenario, NV centers serve as fluorescent reporters of the underlying nitrogen distribution, with clustering directly tracing dopant aggregation. In the latter, even a uniform nitrogen landscape could produce correlated NVs if vacancy migration is influenced by local strain, electrostatic fields, or extended defects that bias the trajectories and recombination locations of mobile vacancies. For example, vacancy traps or scattering centers such as dislocations or ion damage could enhance local capture rates, while strain gradients may create preferential diffusion paths.

If nitrogen clustering is the dominant factor, several mechanisms may be responsible: (*i*) dopant-rich regions arising from step-edge incorporation during CVD growth[27], consistent with terrace widths of 100–300 nm; (*ii*) nitrogen segregation near dislocations or sector boundaries[28-31]; and (*iii*) vacancy-mediated aggregation into A- or B-center complexes prior to NV formation. Recent high-field EPR studies provide direct evidence for spin-active nitrogen clusters and dopant aggregation in type 1b diamond, lending support to this possibility[32,33].

Lastly, we note that a third scenario — where the presence of one NV center directly stimulates the formation of another nearby, through strain or electrostatic interaction — can largely be ruled out. Such "templating" effects would act over only a few nanometers[34], far shorter than the ~100 nm clustering scale observed here.

Beyond constraining possible mechanisms, the present measurements also highlight how wide-field PLE imaging can be used to systematically investigate the origins of NV clustering. Because the technique enables parallel identification and localization of hundreds of individual emitters over large fields of view, it is well suited to comparative studies in which growth, implantation, and annealing conditions are varied independently. For example, comparing samples grown under different nitrogen incorporation conditions but subjected to similar vacancy-generation and annealing protocols could reveal whether clustering tracks dopant inhomogeneity established during CVD growth. Conversely, samples with comparable nitrogen concentrations but different implantation energies, damage profiles, or annealing temperatures could test whether spatial correlations arise primarily from biased vacancy diffusion and capture. Correlative measurements combining wide-field PLE with structural probes such as cathodoluminescence, strain mapping, or electron microscopy could further determine whether clustered NVs preferentially occur near dislocations, sector boundaries, or other extended defects. Such experiments should make it possible to disentangle the relative roles of dopant inhomogeneity and vacancy-mediated processes in shaping the spatial statistics of NV formation.

## 3. CONCLUSIONS

In sum, we have demonstrated a high-throughput, cryogenic PLE imaging platform capable of resolving and characterizing hundreds of NV centers in CVD-grown diamond with sub-diffraction spatial resolution. By analyzing spatial point patterns across wide fields of view, we uncover clear statistical deviations from randomness, including a pronounced tendency for NV centers to form tightly spaced clusters. Through comparison with synthetic control datasets, we show that these correlations are unlikely to arise from stochastic processes alone, suggesting that NV formation is influenced by spatially correlated mechanisms — potentially linked to local strain fields, impurity gradients, or defect-defect interactions during growth.

These findings carry both practical and conceptual implications. On the materials side, the presence of NV clustering offers new insight into the microscopic dynamics of defect formation in diamond, highlighting the need for growth protocols that either suppress or harness correlated behavior, depending on the target application. On the quantum technology front, naturally occurring NV clusters constitute an underexplored resource in the form of small ensembles of proximal spins with potentially coherent interactions that could be leveraged for multi-qubit operations, enhanced sensing via entanglement[6], or structural analysis via correlated electric noise[7,8]. As spatially resolved spectroscopic techniques continue to mature, our approach offers a scalable path toward identifying and monitoring the formation of such functional quantum structures in situ. More broadly, it suggests a shift in how we study and utilize color centers — not merely as isolated emitters, but as collective entities whose spatial and statistical properties can inform materials science and enable new modalities in quantum sensing and information processing.

## DATA AVAILABILITY

The data that support the findings of this study are available from the corresponding author upon reasonable request.

## CODE AVAILABILITY

All source codes for data analysis and numerical modeling used in this study are available from the corresponding author upon reasonable request.

## ACKNOWLEDGMENTS

J.S. and C.A.M. acknowledge support from the U.S. National Science Foundation via grants NSF-2216838 and NSF-2514938, respectively. R.M. acknowledges NSF support via grant NSF-2316693. T.D. acknowledges support by the U.S. Department of Energy, Office of Science, National Quantum Information Science Research Centers, Co-design Center for Quantum Advantage (C2QA) under contract number DE-SC0012704. All authors acknowledge access to the facilities and research infrastructure of the NSF CREST IDEALS, grant number NSF-2112550.

## APPENDIX A: EXPERIMENTAL DETAILS

### 1. Optical microscopy

All measurements were performed in a closed-cycle cryogenic microscope system operated at 9 K under high vacuum. A long-working-distance objective lens (numerical aperture 0.75, vacuum compatible) was used to both focus excitation light and collect photoluminescence (PL) from diamond samples mounted on piezoelectric nano-positioners. Emission collected by the objective was routed via a magnetic-mount dichroic mirror placed outside the cryostat, allowing flexible redirection to either an avalanche photodiode (APD) or an electron-multiplying CCD camera (EMCCD, Princeton Instruments ProEM-HS:512BX3) without altering the excitation pathway. In practice, the EMCCD is operated without threshold, outside of the photon counting regime; the latter allows us to retain a high dynamical range in PL measurement across the collection of optical resonances, likely at the expense of a partial signal-to-noise ratio (SNR) loss if the parameters were optimized for a specific optical resonance.

Excitation was provided by two independent laser sources: a 532 nm green laser (1–5 mW) for NV charge-state initialization and a tunable, narrow-linewidth red laser (637 nm, 0.05–20 μW) for resonant photoluminescence excitation (PLE) and charge readout. Acousto-optic modulators controlled each beam, and an additional shutter was used on the green laser to eliminate leakage during resonant scans. A high-precision wavemeter (High Finesse) continuously monitored the resonant laser wavelength to ensure frequency stability. Measured powers were attenuated by a factor of 3.5 before reaching the objective's back aperture. Microwave excitation was delivered via a 25 μm-diameter copper wire positioned directly on the diamond surface. This provided continuous-wave driving to repump NV centers and prevent shelving into optically dark states during resonant excitation. All experiments were carried out in the absence of externally applied magnetic fields.

We use two alternative methods to achieve wide-field illumination: we either focus the excitation beam onto the microscope objective to create a wide excitation beam at the objective focal plane, or exploit the reflection from the diamond back-side[23]. The back-reflection illumination capitalizes on the diamond's internal reflection at the back-surface: by underfilling the objective's back aperture we obtain a weakly focused beam that forms a broad excitation profile (~75–150 μm diameter) at the focal plane after propagating back and forth through the 200 μm-thick sample. Emission was filtered using a 532 nm notch filter and a 650 nm long-pass filter before being imaged onto the CCD with a 25 cm focal length lens, yielding a total system magnification of 173×.

### 2. Diamond crystals

All data were acquired using electronic-grade, CVD-grown diamond substrates sourced from Delaware Diamond Knives. In both samples, NV formation was stimulated by migration of vacancies during annealing. Specifically, the diamonds were implanted with nitrogen (2.5-10 keV $^{15}N$ for sample A, and 20 MeV $^{14}N$ with a focused ion beam for sample B) and subsequently annealed up to 850 ºC[25] and 1200 ºC[24] for sample A and B respectively. In this study, we primarily analyze NV centers arising from native nitrogen activation, away from implanted regions, which were only used as internal references. Aside from the higher NV density in implanted areas, no systematic differences in optical or spectral behavior were observed between in-grown and bulk implanted NVs under the experimental conditions used, which is consistent with implantation damage having been repaired or not affecting optical resonances[35].

## APPENDIX B: DATA COLLECTION AND PROCESSING

To enable parallel interrogation of many individual NV centers, we employed a wide-field PLE imaging protocol under cryogenic conditions. A green initialization pulse (532 nm) was first delivered in wide-field mode to prepare NV centers into their negative charge state across a broad area of the sample. This was followed by resonant excitation using a frequency-tunable red laser (637 nm), also applied in wide-field geometry, which was scanned across the zero-phonon line to probe the optical transitions of each emitter. The resulting emission, predominantly from phonon sidebands, was collected using an EMCCD camera, enabling us to construct frequency-resolved PL maps over fields of view spanning 48×48 $\mu m^2$.

Each image in the dataset corresponds to a specific laser frequency and contains spatially resolved PL signals from multiple NVs. Finding the maximum for each pixel over the full frequency range produces a composite intensity map[3] in which each bright spot marks the presence of an individual emitter or a sub-diffraction cluster. By analyzing the frequency-resolved stacks, we can resolve and track the spectral features associated with each NV, including their charge-state dynamics under resonant excitation.

Raw data were processed using a custom multi-stage pipeline designed to enhance signal-to-noise ratio and extract spectrally and spatially resolved features from the imaging stack. A three-dimensional Gaussian filter was first applied to smooth noise across both the spatial $(x,y)$ and frequency axes. The kernel widths were chosen to match the characteristic extent of the signal features in the data: approximately 400 nm in real space, corresponding to the typical lateral point-spread function of individual NV emitters in our imaging configuration, and about 100 MHz in frequency, corresponding to the typical linewidth of the optical transitions under our experimental conditions[36]. These values were determined by fitting isolated resonances in the unfiltered dataset and selecting kernel widths that optimize the detection of such peaks while preserving their spatial and spectral localization. Choosing kernel widths comparable to the intrinsic signal dimensions improves the signal-to-noise ratio without artificially broadening or merging nearby features; significantly smaller kernels reduce noise suppression, while excessively large kernels can smear adjacent resonances and increase the probability of false positives or negatives. The smoothed data were then binarized using an adaptive thresholding scheme, which accounts for local variations in background fluorescence and effectively isolates resonance features. Detected features were labeled as "blobs" and further analyzed based on their morphology and spectral properties. For each surviving feature, we extracted the central frequency, linewidth, and lateral position.

To distinguish overlapping emitters and assign resonances to individual NV centers, we leveraged repeated measurements across time. Optical transitions belonging to the same NV tend to exhibit correlated charge dynamics, including synchronous ionization and recovery under strong resonant excitation. To quantify these observations, we compared the likelihood that each pair of optical resonances originates from one or two emitters. The ratio of these likelihoods (Bayes factor) was calculated independently using positional information and blinking. The product of the two Bayes factors was finally used to generate an association matrix, allowing us to group resonances with high confidence even in cases where spatial resolution alone was insufficient to resolve emitters. Each group contains between one and five resonances, mostly under four, which is consistent with each NV possessing six lines, two of them overlapping and four of them having low emitted PL due to poor cyclicity.

This analysis framework enables robust identification and localization of individual NV centers, including those in closely spaced clusters, and forms the basis for our subsequent statistical investigation of emitter distributions. This method functions well until about ten NVs per diffraction-limited spots, at which density the overlap between optical resonances from different NVs cannot be overcome without the use of additional resources, such as for example the application of a magnetic field to lift the degeneracies between the magnetic resonances of differently oriented NVs.

## 1. Distance Based Metric

The distance-based probability metric is a Bayesian estimation of pair-wise likelihood between two optical resonances based on the following mutually exclusive hypotheses, namely,

$H_1$: The optical resonances are different transitions of the same emitter,

$H_2$: The optical resonances stem from two distinct emitters.

$p(H_1)$ quickly becomes negligible as the distance between the resonances exceeds the diffraction limit, hence making calculations necessary only for optical resonance pairs that are in proximity.

The conditional probability of $H_1$ given the displacement vector between the two measured two optical resonances $\boldsymbol{\delta r} = (\delta x, \delta y)$ is formulated using Bayes' theorem

$$P(H_1|\boldsymbol{\delta r}) = \frac{P(\boldsymbol{\delta r}|H_1)P(H_1)}{P(\boldsymbol{\delta r})}\,. \tag{1}$$

Since there are only two possible hypotheses, the total probability must add to one

$$P(H_1|\boldsymbol{\delta r}) + P(H_2|\boldsymbol{\delta r}) = 1\,, \tag{2}$$

such that

$$P(H_1|\boldsymbol{\delta r}) = c \cdot P(\boldsymbol{\delta r}|H_1)P(H_1)\,, \tag{3}$$

$$P(H_2|\boldsymbol{\delta r}) = c \cdot P(\boldsymbol{\delta r}|H_2)P(H_2)\,, \tag{4}$$

where $c$ is a constant of proportionality, and in this case equivalent to $1 / P(\boldsymbol{\delta r})$. Using Eqs. (2-4) to solve for $c$, we find

$$c = \frac{1}{P(H_1)P(\boldsymbol{\delta r}|H_1) + P(H_2)P(\boldsymbol{\delta r}|H_2)}\,. \tag{5}$$

By substituting $c$ in place of $1 / P(\boldsymbol{\delta r})$ in Eq. (1), the Bayes' theorem now takes the following form

$$p(H_1|\boldsymbol{\delta r}) = \frac{p(\boldsymbol{\delta r}|H_1) \cdot p(H_1)}{p(\boldsymbol{\delta r}|H_1)p(H_1) + p(\boldsymbol{\delta r}|H_2)p(H_2)}\,. \tag{6}$$

We expand the term involving $H_2$ into integral form, with respect to the actual displacement $\boldsymbol{\delta r'}$

$$p(\boldsymbol{\delta r}|H_2)p(H_2) = \int p(\boldsymbol{\delta r}|H_2, \boldsymbol{\delta r'})p(H_2, \boldsymbol{\delta r'})d^2\boldsymbol{\delta r'}, \tag{7}$$

where $p(H_2, \boldsymbol{\delta r'})d^2\boldsymbol{\delta r'}$ is the probability of having two emitters that are separated by $\boldsymbol{\delta r'}$. It is well approximated as $p(H_1)\rho_{\mathrm{NV}}d^2\boldsymbol{\delta r'}$, where $\rho_{\mathrm{NV}}$ is the NV density. The probability distribution to measure $\boldsymbol{\delta r} = (\delta x, \delta y)$ given two NVs that are separated by $\boldsymbol{\delta r'} = (\delta x', \delta y')$ can be written as

$$p(\boldsymbol{\delta r}|H_2, \boldsymbol{\delta r'}) = G(\delta x - \delta x', \sigma_x).G(\delta y - \delta y', \sigma_y), \tag{8}$$

where we use normal distributions, $G(u, \sigma_u) = \frac{1}{\sqrt{2\pi}\sigma_u}\exp\left[\frac{-\delta u^2}{2\sigma_u^2}\right]$, along each of the dimensions and consider the estimated errors $\sigma_x$ and $\sigma_y$ for the measurement of the optical resonance displacement $\boldsymbol{\delta r}$ based on the standard deviation of repeated measurements. Since it takes the form of a Gaussian, it integrates to 1. With the above substitutions and simplifications, we can expand the integral in the two spatial dimensions to get

$$\begin{aligned}\int p(\delta r|H_2, \boldsymbol{\delta r'})p(H_2, \boldsymbol{\delta r'})d\boldsymbol{\delta r'} &= \\ \iint p(H_1)\rho_{\mathrm{NV}}G(\delta x - \delta x', \sigma_x)G(\delta y - \delta y', \sigma_y)d\delta x' d\delta y' \\ &= p(H_1)\rho_{\mathrm{NV}}\,.\end{aligned} \tag{9}$$

This then further simplifies the initial probability to

$$p(H_1|\boldsymbol{\delta r}) = \frac{p(\boldsymbol{\delta r}|H_1)}{p(\boldsymbol{\delta r}|H_1) + \rho_{\mathrm{NV}}} = \frac{1}{1 + \frac{\rho_{\mathrm{NV}}}{p(\boldsymbol{\delta r}|H_1)}}\,. \tag{10}$$

Finally, we can estimate $p(\boldsymbol{\delta r}|H_1) = G(\delta x, \sigma_x).G(\delta y, \sigma_y)$ based on the error of the displacement measurement.

## 2. Time Dynamic Based Metric

Our blinking metric for two optical resonances $A$ and $B$ repetitively measured is based on an estimation of the ratio of probabilities to have one or two emitters given our measurement result (Bayes factor). We obtain the formula for the Bayes factor by first estimating the probability of carrying a measurement under the two different hypotheses, $H_1$ and $H_2$.

Under $H_1$ (single emitter), for each spectra the probability to observe a resonance $X$ ($A$ or $B$) is $P_{X,1} = \pi \cdot p_X + (1 - \pi) \cdot f_X$ where $\pi$ is the probability for the emitter to be bright, $p_X$ is the detection probability when the emitter is bright, and $f_X$ the probability of a false positive. For each spectrum, the probability for four observables $P_{i_A i_B}$ — where $i_X$ is 1 if $X$ is detected, 0 otherwise — can then be written:

$$P_{11} = \pi p_A p_B + (1 - \pi)f_A f_B\,, \tag{11}$$

$$P_{10} = \pi p_A(1 - p_B) + (1 - \pi)f_A(1 - f_B)\,, \tag{12}$$

$$P_{01} = \pi(1 - p_A)p_B + (1 - \pi)(1 - f_A)f_B\,, \tag{13}$$

$$P_{00} = \pi(1 - p_A)(1 - p_B) + (1 - \pi)(1 - f_A)(1 - f_B). \tag{14}$$

Under $H_1$ the result of a series of spectra therefore has a probability $L_1 = P_{11}^{n_{11}}.P_{10}^{n_{10}}.P_{01}^{n_{01}}.P_{00}^{n_{00}}$, where $n_{i_A i_B}$ is the number of times the pair of resonances $A$ and $B$ were in the state $i_A, i_B$.

Similarly, under $H_2$ (two emitters), we can calculate the probability of each of the observables. Defining $q_X$ as the probability to measure an optical resonance $X$ (including a false positive), we find a probability for a series of spectra under $H_2$ of $L_2 = (q_A q_B)^{n_{11}}[q_A(1 - q_B)]^{n_{10}}[(1 - q_A)q_B]^{n_{01}}[(1 - q_A)(1 - q_B)]^{n_{00}}$.

Using the Bayes theorem, and assuming no prior, we finally find a Bayes factor for a given measurement $M$:

$$\mathcal{B}(M) = P(H_1|M)/P(H_2|M) = L_1/L_2\,. \tag{15}$$

In practice, we estimate $\pi p_X$ and $q_X$ for $X \in \{A, B\}$ from the detection statistics of the spectra series, neglecting false positives.

### 3. Merging

After calculating all pairwise metrics, we are left with a $n \times n$ symmetric matrix $\boldsymbol{\mathcal{B}}^{\mathbf{0}}$ , where $n$ is the number of optical resonances, and matrix element $\mathcal{B}^{0}_{ij}$ is the pairwise metric Bayes' factor between resonance $i$ and resonance $j$. To infer from $\boldsymbol{\mathcal{B}}^{\mathbf{0}}$ the correct optical resonance groupings based on their source NVs, we use an iterative algorithm that combines the Bayes factor for optical resonances that are the most likely to belong to the same emitter.

The algorithm begins by merging optical resonances with the highest Bayes factor and continues until no Bayes' factor greater than one remains. The $m$-th merger uses the Bayes factor in the matrix $\boldsymbol{\mathcal{B}}^{\boldsymbol{m-1}}$ of size $n+1-m$ and leads to the matrix $\boldsymbol{\mathcal{B}}^{\boldsymbol{m}}$ of size $n-m$ . To merge resonances $i < j$, we combine their Bayes factors multiplicatively:

$$\mathcal{B}^{m}_{i,1\ldots n-m} = {\mathcal{B}^{m}_{1\ldots n-m,i}}^{T} = \mathcal{B}^{m-1}_{i,1\ldots n+1-m} \cdot \mathcal{B}^{m-1}_{j,1\ldots n+1-m}\,, \quad (16)$$

while deleting the $j$-th row and column of $\boldsymbol{\mathcal{B}}^{\boldsymbol{m}}$ and keeping all other Bayes factors as in $\boldsymbol{\mathcal{B}}^{\boldsymbol{m-1}}$.

For each iteration, this method consists in assuming that the two optical resonances most likely to belong to the same emitter do belong to the same emitter, allowing us to combine the information gathered with respect to other resonances, thereby improving the signal to noise ratio for more ambiguous pairs.

### 4. Drift Correction

Once all well resolved blobs have been extracted from the data, they are then projected over time, so that all measured instances are placed in a shared spatial and frequency space. Blobs that belong to the same optical resonance form clusters in this shared space as repeated measurements of a given optical resonance should yield blobs in the same position. These clusters are assigned using the DBSCAN (Density-Based Spatial Clustering of Applications with Noise) algorithm. The neighborhood search radius is set based on statistical variations between repeated measurements and scaled appropriately for each individual dimension. Sometimes, the variations across measurements result in clusters that are more spread out, which interferes with the clustering algorithm. Most notably, sample drift can cause significant deviations in position that must be corrected before applying DBSCAN. Since the change in position due to sample drift should be homogeneous for all optical resonances, we manually select the blobs of an isolated optical resonance with high signal-to-noise ratio for drift characterization. This is done by fitting blob displacement over time using a polynomial function. Then the drift is subtracted from each blob accordingly. The uncertainty of this optical resonance's position (typically about 0.5 pixel) is added to all others to account for the additional errors that could have accrued from the drift correction. Note that with further improvements in drift correction and SNR, the precision of position measurements could eventually reach lattice-site precision as recently highlighted in Ref. [37].